\documentclass[aps,prx,longbibliography,twocolumn,superscriptaddress]{revtex4-2}
\usepackage{graphicx}
\usepackage{color}
\usepackage{multirow}
\usepackage{amsmath}
\setcitestyle{numbers,square,citesep={,\kern-.30em}} %for making no space between number in the citation box.

\begin{document}

% Use the \preprint command to place your local institutional report
% number in the upper righthand corner of the title page in preprint mode.
% Multiple \preprint commands are allowed.
% Use the 'preprintnumbers' class option to override journal defaults
% to display numbers if necessary
%\preprint{}

%Title of paper
\title{Estimations of On-Site Coulomb Potential and Covalent State in La$_2$CuO$_4$ by Muon Spin Rotation and Density Functional Theory Calculation }

\author{Muhammad Redo Ramadhan}
\email{corresponding author; redo.ramadhan@idu.ac.id; present address: Department of Chemical Engineering, Faculty of Industrial Technology, Universitas Pembangunan Nasional "Veteran" Yogyakarta,
Sleman, Yogyakarta 55283, Indonesia.}
\affiliation{Meson Science Laboratory, RIKEN Nishina Center, 2-1 Hirosawa, Wako, Saitama 351-0198, Japan,}
\affiliation{Department of Physics, Universitas Indonesia, Depok 16424, Indonesia,}
\author{Budi Adiperdana}
\affiliation{Meson Science Laboratory, RIKEN Nishina Center, 2-1 Hirosawa, Wako, Saitama 351-0198, Japan,}
\affiliation{Department of Physics, Universitas Padjajaran, Sumedang 45363, Indonesia,}
\author{Irwan Ramli}
\email{present address: Department of Physics, Universitas Cokroaminoto Palopo, Kota Palopo 91911, Indonesia.}
\affiliation{Meson Science Laboratory, RIKEN Nishina Center, 2-1 Hirosawa, Wako, Saitama 351-0198, Japan,}
\affiliation{Department of Condensed Matter Physics, Hokkaido University, Sapporo 060-8010, Japan,}
\author{Dita Puspita Sari}
\email{present address: Graduate School of Engineering and Science, Shibaura Institute of Technology, Saitama 337-8570, Japan.}
\affiliation{Meson Science Laboratory, RIKEN Nishina Center, 2-1 Hirosawa, Wako, Saitama 351-0198, Japan,}
\affiliation{Department of Physics, Osaka University, Osaka 560-0043, Japan,}
\author{Anita Eka Putri}
\affiliation{Meson Science Laboratory, RIKEN Nishina Center, 2-1 Hirosawa, Wako, Saitama 351-0198, Japan,}
\affiliation{Department of Physics, Universitas Indonesia, Depok 16424, Indonesia,}
\author{Utami Wydiaiswari}
\affiliation{Meson Science Laboratory, RIKEN Nishina Center, 2-1 Hirosawa, Wako, Saitama 351-0198, Japan,}
\affiliation{Department of Physics, Universitas Indonesia, Depok 16424, Indonesia,}
\affiliation{Department of Condensed Matter Physics, Hokkaido University, Sapporo 060-8010, Japan,}
\author{Harion Rozak}
\email{joint address: USM-RIKEN Interdisciplinary Collaboration for Advanced Sciences, School of Distance Education, Universiti Sains Malaysia, 11800 Minden, Pulau Pinang, Malaysia; present address: Graduate School of Engineering and Science, Shibaura Institute of Technology, Saitama 337-8570, Japan.}
\affiliation{Meson Science Laboratory, RIKEN Nishina Center, 2-1 Hirosawa, Wako, Saitama 351-0198, Japan,}
\affiliation{Computational Chemistry and Physics Laboratory, School of Distance Education, Universiti Sains Malaysia, Pulau Pinang 11800, Malaysia,}
\author{Wan Nurfadhilah Zaharim}
\email{joint address: USM-RIKEN Interdisciplinary Collaboration for Advanced Sciences, School of Distance Education, Universiti Sains Malaysia, 11800 Minden, Pulau Pinang, Malaysia.}
\affiliation{Meson Science Laboratory, RIKEN Nishina Center, 2-1 Hirosawa, Wako, Saitama 351-0198, Japan,}
\affiliation{Computational Chemistry and Physics Laboratory, School of Distance Education, Universiti Sains Malaysia, Pulau Pinang 11800, Malaysia,}
\author{Azwar Manaf}
\affiliation{Department of Physics, Universitas Indonesia, Depok 16424, Indonesia,}
\author{Budhy Kurniawan}
\email{corresponding author; budhy.kurniawan@sci.ui.ac.id.}
\affiliation{Department of Physics, Universitas Indonesia, Depok 16424, Indonesia,}
\author{Mohamed Ismail Mohamed-Ibrahim}
\affiliation{Computational Chemistry and Physics Laboratory, School of Distance Education, Universiti Sains Malaysia, Pulau Pinang 11800, Malaysia,}
\author{Shukri Sulaiman}
\email{joint address: USM-RIKEN Interdisciplinary Collaboration for Advanced Sciences, School of Distance Education, Universiti Sains Malaysia, 11800 Minden, Pulau Pinang, Malaysia.}
\affiliation{Meson Science Laboratory, RIKEN Nishina Center, 2-1 Hirosawa, Wako, Saitama 351-0198, Japan,}
\affiliation{Computational Chemistry and Physics Laboratory, School of Distance Education, Universiti Sains Malaysia, Pulau Pinang 11800, Malaysia,}
\author{Takayuki Kawamata}
\email{present address: Department of Natural Sciences, Tokyo Denki University, Tokyo 120-8551, Japan.}
\affiliation{Department of Applied Physics, Tohoku University, Sendai 980-8579, Japan,}
\author{Tadashi Adachi}
\affiliation{Department of Engineering and Applied Sciences, Sophia University, Tokyo 102-8554, Japan.}
\author{Yoji Koike}
\affiliation{Department of Applied Physics, Tohoku University, Sendai 980-8579, Japan,}
\author{Isao Watanabe}
\email{corresponding author; nabedon@riken.jp}
\affiliation{Meson Science Laboratory, RIKEN Nishina Center, 2-1 Hirosawa, Wako, Saitama 351-0198, Japan,}
\affiliation{Department of Physics, Universitas Indonesia, Depok 16424, Indonesia,}
\affiliation{Department of Physics, Universitas Padjajaran, Sumedang 45363, Indonesia,}
\affiliation{Department of Condensed Matter Physics, Hokkaido University, Sapporo 060-8010, Japan,}
\affiliation{Department of Physics, Osaka University, Osaka 560-0043, Japan,}
\affiliation{Computational Chemistry and Physics Laboratory, School of Distance Education, Universiti Sains Malaysia, Pulau Pinang 11800, Malaysia,}
\date{\today}

\begin{abstract}
The on-site Coulomb potential, $U$, and the covalent state of electronic orbitals play key roles for the Cooper pair symmetry and exotic electromagnetic properties of high-$T_{\rm c}$ superconducting cuprates. In this report, we demonstrate a new way to determine the value of $U$ and present the whole picture of the covalent state of Cu spins in the mother system of the La-based high-$T_{\rm c}$ superconducting cuprate, La$_2$CuO$_4$, by combining the muon spin rotation ($\mu$SR) and the density functional theory (DFT) calculation. We succeeded in revealing local deformations of CuO$_6$ octahedron followed by changes in Cu-spin distributions which were caused by the injected muon. Adjusting the DFT and $\mu$SR results, $U$ and the minimum charge transfer energy between the upper Hubbard band and the O2$p$ band were optimized to be 4.87(4) eV and 1.24(1) eV, respectively.  
\end{abstract}

%\keywords{}
%\maketitle must follow title, authors, abstract, and keywords
\maketitle{}

\section{introduction}
The La-based high-$T_{\rm c}$ superconducting cuprate is a typical Mott system and has a rich variety in physics, making this system still mysterious and brightly fascinating even after three decades have passed since its discovery. There are open questions on exotic electronic states that need to be investigated, like pseudogaps,\cite{Timusk} stripes of spins and holes,\cite{Tranquada} precursor of superconduting states,\cite{Wang} unconventional normal states\cite{Boebinger} and charge-ordered states.\cite{Chang} These unique states are commonly realized on the basis of the strong on-site Coulomb potential, $U$, and covalent states of Cu3$d$ orbitals with surrounding O2$p$ orbitals.\cite{Czyzyk,Anisimov0,Wan,Pesant,Werner,Jang,Hirayama,Lane,Nilsson} Both properties have been suggested to be essential to describe the possible mechanism of the high-$T_{\rm c}$ superconductivity because those carry the symmetry of the wave function of the Cooper pair and electronic conducting properties.\cite{Zhang,Ogata} 

For deeper understanding of those exotic effects caused by the on-site Coulomb potential on Cu, $U$, and the covalent state, the mother system of the La-based high-$T_{\rm c}$ superconducting cuprate, La$_2$CuO$_4$ (LCO), can provide an ideal playground. LCO is a typical Mott insulator and has the strong covalent state of Cu3$d$ with O2$p$ within the two-dimensional (2D) CuO$_2$ plane. The antiferromagnetic (AF) interaction between Cu spins leads to the formation of the AF ordered state.\cite{Budnick,Vaknin,Uemura,Borsa} The exchange coupling energy within the CuO$_2$ plane was suggested to be about 140 meV.\cite{Coldea} The value of $U$ on the Cu atom has been well investigated but still has large ambiguities of 3-10 eV,\cite{Czyzyk,Anisimov0,Wan,Pesant,Werner,Jang,Hirayama,Lane,Nilsson} giving uncertainty on discussions of exotic electronic states of high-$T_{\rm c}$ superconducting cuprates. This is because, those features contain quantum and multi-body effects of electrons which are still difficult to approach either experimentally and theoretically. 

Following this situation, we suggest a novel approach to this problem by combining the muon spin rotation ($\mu$SR) measurement with the density functional theory (DFT) calculation including $U$ as an adjustable parameter (DFT+$U$).\cite{Anisimov1,Dudarev,Varignon} The muon is a sensitive local magnetic probe and can trace the covalent state with helps of DFT+$U$. In this report, we are going to show the results of this combined investigation on LCO, revealing the covalent state of Cu spins and determining $U$. We found three muon sites in LCO. Those muon positions were described from our DFT+$U$ with the full view of the spatial distribution of Cu spins caused by the covalent state. Adjusting DFT+$U$ with the $\mu$SR results, we obtained the $U$ value to be 4.87(4) eV followed by the determination of the minimum charge-transfer (CT) energy between the upper Hubbard band of Cu3$d_{x^2-y^2}$ and O2$p$ to be 1.24(1) eV, and the size of the magnetic moment of Cu spin to be 0.520(3) $\mu_B$. 

\section{Experimentals}
\subsection{Growth of the La$_2$CuO$_4$ single crystal}
A large LCO single-crystal was synthesized by the traveling-solvent floating-zone method and was confirmed to be of a single phase without impurities by using the X-ray diffraction measurement at room temperature. After the oxygen reduction annealing in Ar-gas flow, the AF transition temperature, $T_{\rm N}$, was estimated from the  magnetic susceptibility measurement by using a Superconducting Quantum Interferometric Device (Quantum Design Co. Ltd., MPMS-XL). The crystal was sliced in parallel with the CuO$_2$ layer for present $\mu$SR measurements.

\subsection{$\mu$SR}
$\mu$SR measurements were carried out on the GPS spectrometer at the Paul Scherrer Institut (PSI) in Switzerland by using a continuous muon source in the zero-field condition.  The muon was injected into the LCO single-crystal sample keeping the initial spin-polarization to be perpendicular to the CuO$_2$ plane. The time dependence of the asymmetry parameter, $A$($t$), is defined as 
$A$($t$) = $\frac{F(t)-B(t)}{F(t)+B(t)}$ ($\mu$SR time spectrum). Here, $F$($t$) and $B$($t$) are numbers of positron counted by the forward and backward counters at $t$, respectively.\cite{Hayano,Uemura0} In order to determine internal fields at muon sites with the higher accuracy, we gathered more than 600 million positrons which were more than 20 times higher than usual cases. 

\subsection{DFT Calculations}
DFT calculations were conducted using Vienna {\it ab-initio} Simulation Package (VASP)\cite{Kresse1,Kresse2} with the Generalized Gradient Approximation Perdew-Wang91 (GGA-PW91) exchange-correlation functional with adjusting $U$ between 2-8 eV.\cite{Perdew1,Dudarev} The Kohn-Sham approach using the projector augmented-waves (PAW) formalism was adopted as implemented in VASP.\cite{Kresse1,Kresse2} It should be noted that DFT results generally depend on the functional.\cite{Kulik} Although GGA+$U$ is not neither ideal nor the best functional to exactly describe electronic state of LCO, this functional is well known to be valid with $U$ in order to describe electronic states of strongly correlated systems.\cite{Czyzyk,Pesant} Since there is no ideal full self-interaction correlated functional even now, we chose GGA+$U$ as the "best possible" functional for the present study as well as other published papers.\cite{Varignon} 

The ground state of a calculation model was achieved by setting the convergence criterion of 1 $\times$ 10$^{-4}$ eV. The relaxation process of all atomic positions was terminated until the magnitude of the force on each atom became less than 0.05 eV/\AA\ following the quasi-Newton algorithm. The crystal structural symmetry was set to be orthorhombic with the $Bmab$ space group. Lattice parameters for the unit cell were set to be $a$ = 5.3568 \AA\, $b$ = 5.4058 \AA\ and $c$ = 13.1432 \AA\ as estimated by the neutron scattering experiment.\cite{Reehuis} 

\begin{figure}[htb]
\includegraphics[width=5cm]{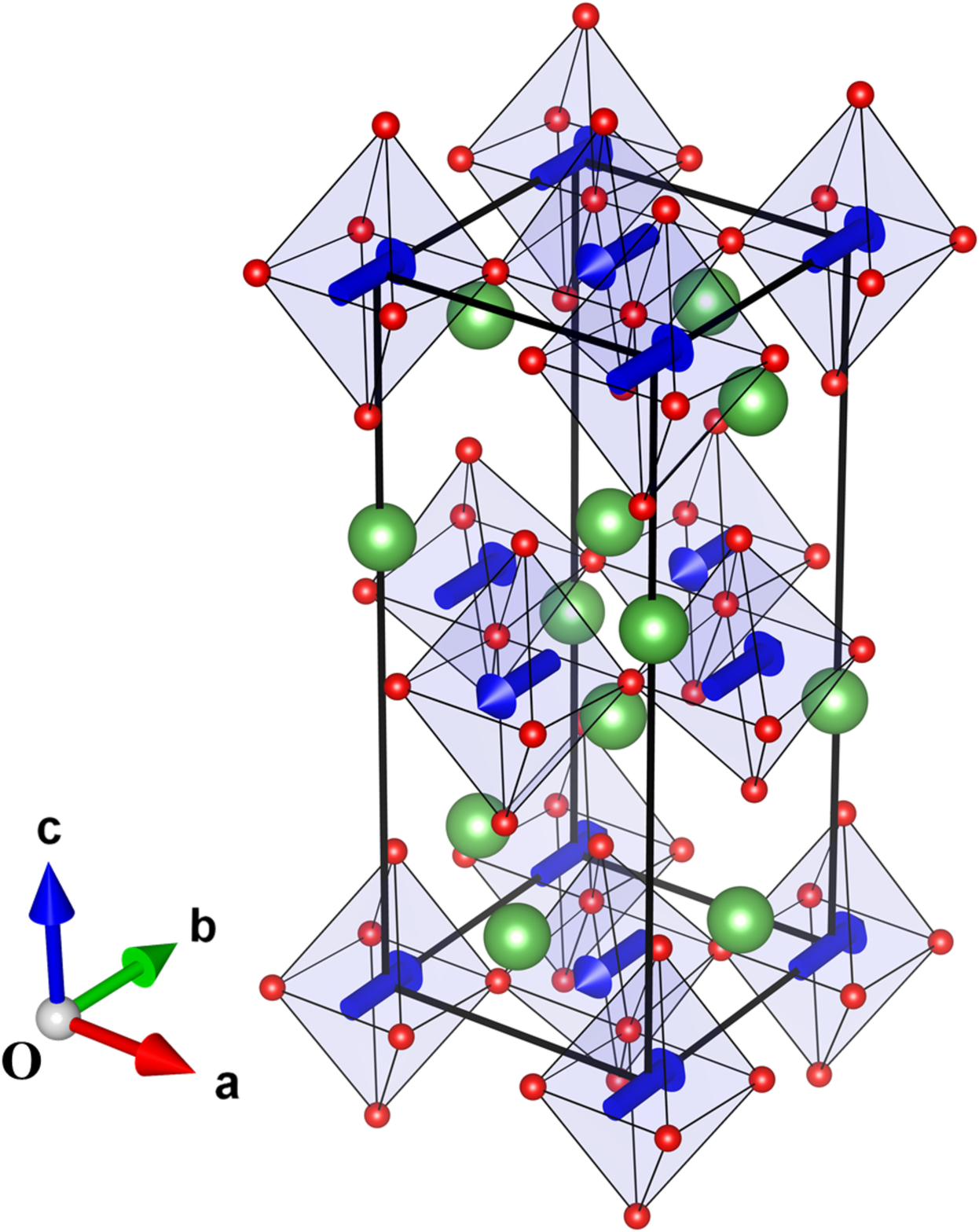}
\caption{\label{spin_structure} Initial condition of the Cu-spin structure for the present DFT calculation study. Blue, red and green marks are Cu spins, O and La atoms, respectively. Cu spins form the AF spin alignment with the spin direction along to the b-axis within the CuO$_2$ plane. This spin structure is the same with that determined from the neutron scattering experiment.\cite{Vaknin}  }
\end{figure}

Figure \ref{spin_structure} indicates an initial condition of the Cu-spin structure for present DFT calculations. Cu spins form the AF alignment with the spin direction in parallel with the b-axis within the CuO$_2$ plane. This spin structure is the same with that determined from the neutron scattering experiment.\cite{Vaknin} The supercell containing 32 unit cells with one muon in the formation of the 4$\times$4$\times$2 stacking was used for all our non-collinear DFT calculations to estimate stable muon positions. RIKEN Supercomputing Facility named HOKUSAI was used for our supercell calculations. 

\section{Results}
\subsection{Characterizations of the La$_2$CuO$_4$ single crystal}
\begin{figure}[htb]
\includegraphics[width=6cm]{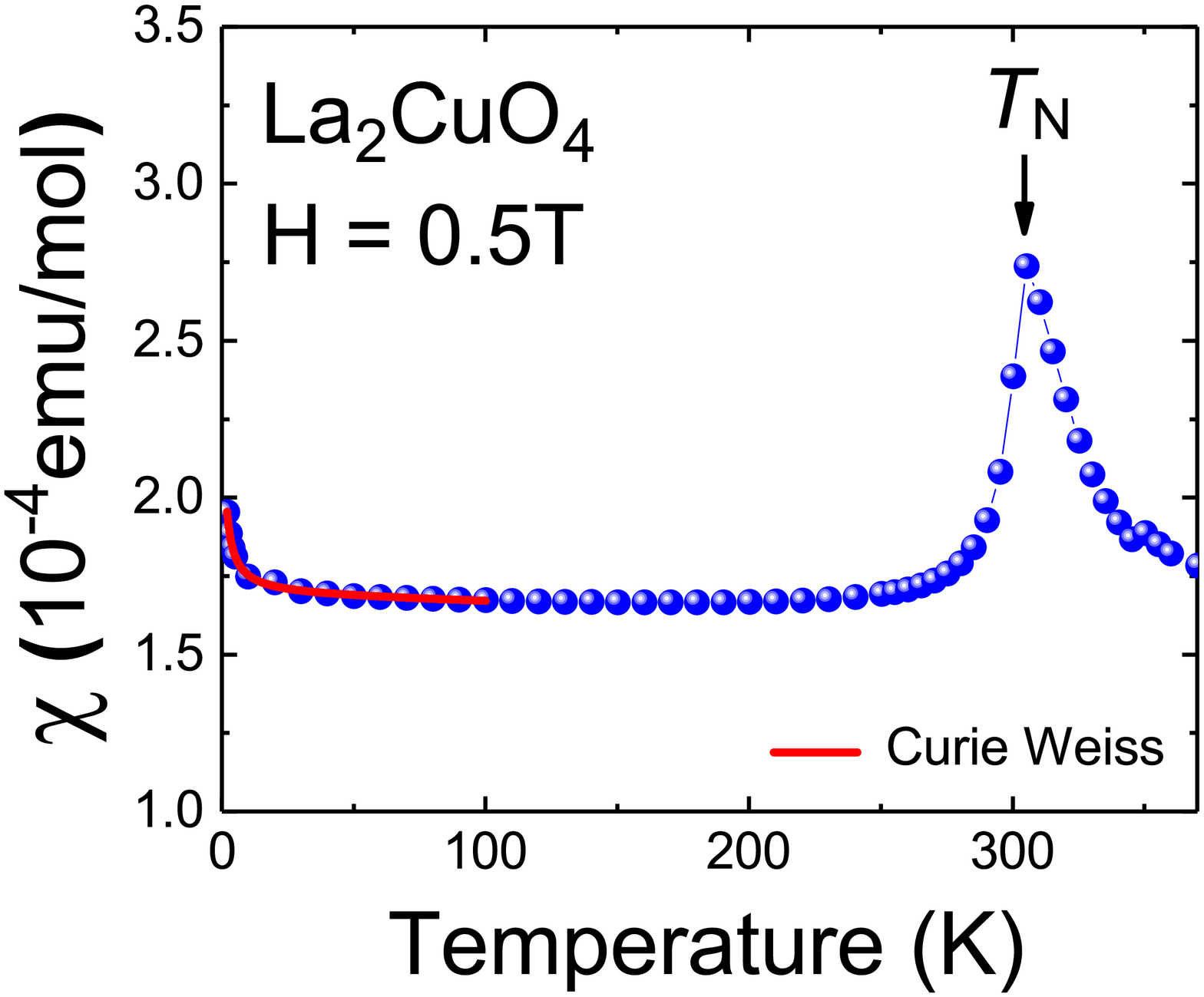}
\caption{\label{susceptibility} Temperature dependence of the magnetic susceptibility of the La$_2$CuO$_4$ single crystal. The magnetic field of 0.5 T was applied perpendicular to the CuO$_2$ plane. The black arrow shows the antiferromagnetic transition temperature, $T_{\rm N}$ of 309 K. The red-solid line indicates the best-fit result by using the Curie-Weiss law below 100 K.  }
\end{figure}

Figure \ref{susceptibility} shows the temperature dependence of the magnetic susceptibility of the LCO single crystal which was used for the present $\mu$SR study. The magnetic field of 0.5 T was applied perpendicular to the CuO$_2$ plane. A sharp peak was observed around 309 K which was due to the appearance of the long-range AF ordering of Cu spins.\cite{Vaknin,Reehuis} An increase in the magnetic susceptibility was observed below about 20 K. This increase was fitted by the Curie-Weiss law. The red-solid line in Fig. \ref{susceptibility} is the best fit result within the temperature range below 100 K. The Weiss temperature was estimated from this low temperature analysis to be  -1.2(2) K. This result indicates that the increase in the magnetic susceptibility below 20 K is due to free spins which are not related to the AF ordering. Assuming that those free spins would be coming from Cu spins which appear around crystal defects, its fraction was estimated to be 0.024 \%. 

\subsection{$\mu$SR}
\begin{figure}[htb]
\includegraphics[width=6cm]{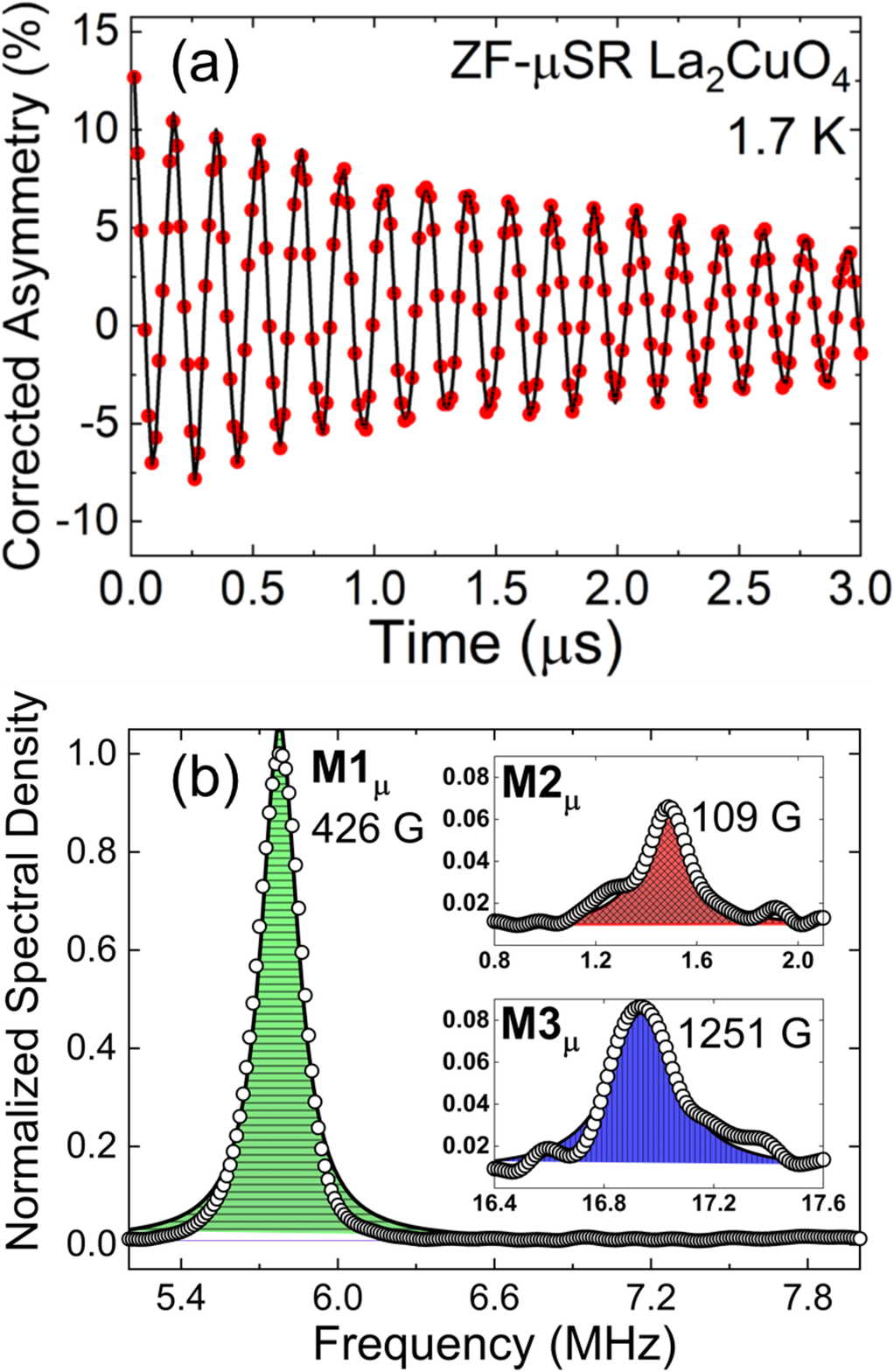}
\caption{(a) $\mu$SR time spectrum measured at 1.7 K on La$_2$CuO$_4$ single crystal. The solid line is the best-fit results by using Eq.(1) with $i$=1,2,3. (b) Fourier spectrum of the muon-spin precession. Solid lines show the best fit results obtained by using the Lorentzian function. Insets show Fourier spectra of additional two components.}
\label{time_spectrum}
\end{figure}

Figure \ref{time_spectrum}(a) shows the $\mu$SR time spectrum measured in the zero-field condition at 1.7 K on the LCO single crystal. The observation of the muon-spin precession proved that Cu spins in LCO were in the AF ordered state.\cite{Uemura} The $\mu$SR time spectrum showed many turns of the muon spin with the slow damping rate. The muon-spin precession was apparent at least up to 6 $\mu$sec which was the reliable maximum measurable time. The observation of the clear muon-spin recession in the long-time region indicates that the AF network of Cu spins is well coherent compared to those used in other $\mu$SR studies.\cite{Uemura,Borsa,PSI}  

Figure \ref{time_spectrum}(b) shows the Fourier spectrum of the muon-spin precession. We confirmed three peaks. One was the main peak with a large spectral weight compared with the other two. The other two peaks were found at the higher and lower frequency sides with much smaller spectral weight than that of the main peak. These results mean that there are three possible muon stopping positions in LCO with different occupancies. For convenience, we named the main peak, lower field and higher field positions as M1$_{\mu}$, M2$_{\mu}$ and M3$_{\mu}$, respectively.

We applied the Lorentzian function to estimate the internal field at each muon site. Solid lines in Fig. \ref{time_spectrum}(b) are the best-fit results for each peak. The frequency value at each peak position,  $\omega$, was converted to an internal magnetic field at the muon site, $H$, by the following relation,
\begin{equation}
\omega = \gamma_\mu H. 
\label{Conversion}
\end{equation}
\noindent
At here, $\gamma_\mu$ is the gyromagnetic ratio of $\mu$ of 135.5 MHz/T. The main peak is corresponding to the internal field of 426.7(1) G ($\sim$5.78 MHz). This value is the same as that reported in the past.\cite{Uemura} The low-frequency one is corresponding to 109.2(4) G ($\sim$1.48 MHz). This low-frequency peak was reported from the $\mu$SR study on the LCO thin film with much bigger muon-precession amplitude compared to our observed one in the bulk LCO single crystal.\cite{PSI} At this moment, reasons why the amplitude obtained in the LCO thin film and the bulk form is different are still unclear and necessary to be investigated. On the other hand, The high-frequency peak with the internal field of 1251.6(3) G ($\sim$17.0 MHz) was not reported in the LCO thin film. There would be a possibility that this high-frequency peak was not clearly observed in the LCO thin film by some reasons, such as too low precession amplitude. 

Following this result, the time spectrum shown in Fig. \ref{time_spectrum}(a) was analyzed assuming three muon sites by applying Eq.(\ref{analysis}). Watching the time spectrum carefully, the center of the muon-spin precession is shifted from the corrected zero-asymmetry position and relax slowly. The shift of the time spectrum from the corrected zero position was included in the analysis function as the offset component. Constant background signals which were coming from surroundings of the sample were subtracted from the time spectrum by applying this fitting method, so that the $\mu$SR time spectrum shown in Figure \ref{time_spectrum}(a) is the background-free spectrum. The solid line in Fig. \ref{time_spectrum}(a) is the best-fit result.
\begin{equation}
A(t)= \sum_i A_i \cos(\omega_i t+\phi_i) e^{-\lambda_i t} + A_{offset} e^{-\lambda_{offset} t}
\label{analysis}
\end{equation}
\noindent
Here, $A_i$ and $A_{offset}$, $\lambda_i$ and $\lambda_{offset}$ are initial asymmetries at $t$=0, relaxing rates of the muon-spin precession and the offset component, respectively. The $\omega_i$ and $\phi_i$ are the frequency and phase of the muon-spin precession, respectively. The $\omega_i$ is converted to the the interal field at the muon site following Eq. \ref{Conversion}. For convenience sake, we indexed the internal field at each muon site to be  $H^{\rm M \it i}_{\mu {\rm SR}}$  ($i$=1,2,3).  The ratio among $A_i$ is corresponding to the existing probability of the muon at each stopping site, putting $i$=1,2,3 for M1$_{\mu}$, M2$_{\mu}$ and M3$_{\mu}$, respectively. All parameters obtained from the Fourier analysis and direct fitting of the time spectrum are listed in Table 1. 

The ratio among initial asymmetries seems to be different from that of the Fourier spectrum weight. This is due to the fast relaxation rate for M2$_{\mu}$ compared to those for M1 and M3. After compensating the Fourier spectrum wight by the relaxation rate, both ratios became similar each other. Accordingly, we used the ratio among initial asymmetries for simplicity to argue the population of the muon at each site.

\onecolumngrid

\begin{center}
\begin{table}[htb]
\fontsize{10pt}{0.5cm}\selectfont
\caption{Obtained parameters from the best-fit of the $\mu$SR time spectrum by using Eq.(\ref{analysis}) and Fourier spectra by using the Gaussian function. The $\omega_i$ was converted to the internal field at each muon site, $H^{\rm M \it i}_{\rm {\mu SR}}$, following Eq. \ref{Conversion}.  } 
\begin{tabular}{c||cccc|c} \hline
            & \multicolumn{4}{c|}{$\mu$SR Time Spectrum} & Fourier Spectra \rule[0mm]{0mm}{3mm}\\ \cline{2-6}
    &$A_i$ (\%) & $H^{\rm M \it i}_{\mu {\rm SR}}$ (G) & $\phi_i$ (degree)& $\lambda_i$ ($\mu$sec$^{-1}$) & Peak Position (G) \rule[0mm]{0mm}{3mm} \\ \hline \hline
M1$_{\tiny \rm {\mu}}$ &9.228(23)&426.35(2)&-5.15(15)&0.333(20)&426.59 (1) \\
M2$_{\tiny \rm {\mu}}$ &2.631(54)&95.9(58)&22.6(55)&5.36(26)&109.16 (39)\\
M3$_{\tiny \rm {\mu}}$ &0.872(26)&1245.54(36)&-5.7(17)&0.50(3)& 1251.55 (27) \\
offset                 &1.879(54)&---&---&0.380(36)&--- \\ \hline
\end{tabular}
\label{Analysis_Results}
\end{table}
\end{center}

\twocolumngrid

\subsection{DFT+$U$ without $\mu$}
In the first attempt to combine DFT+$U$ with the $\mu$SR results, we tried to visualize the full view of the covalent state of Cu around the the nuclear position without the muon. Figure \ref{Cu_spin}(a) exhibits the map of the Cu-spin density distribution obtained from DFT+$U$ calculations. The $U$ was simply set to be 5.0 eV as a convenience. The Cu-spin density on the CuO$_2$ plane expands from the atomic position of Cu to the in-plane O sites. This is due to the covalent state of Cu3$d_{x^2-y^2}$ with the neighboring O2$p_{\sigma}$ and causes the appearance of a partial density of Cu-spin at the in-plane O position. The adjacent Cu spin also expands its density to the same in-plane O site with the opposite sign of the spin direction and cancels the net magnetic moment at in-plane O. We found that about 18\% of the Cu spin was transferred from the atomic position of Cu to in-plane O. This should be one of reasons why the net magnetic moment of the Cu-spin ($S$=1/2) is not 1 $\mu_B$ but reduced to be about a half as observed by neutron scattering experiments.\cite{Vaknin,Reehuis} This reduction due to the covalent state explained only 36 \% of the total reduction in the magnetic moment of Cu, indicating that the quantum spin fluctuation effect is important to satisfy the difference.\cite{Kojima} 

In addition to this, a small amount of the asymmetric Cu-spin density was found in 2$p_z$ of apical O as indicated in Fig. \ref{Cu_spin}(a). The estimated amount was at most 1\% of the Cu spin, and the density inside the CuO$_6$ octahedron is bigger than that of the outside. This result is consistent with that obtained by Lane {\it et al.}\cite{Lane} Consequently, the net magnetic moment at apical O is not canceled and the amount of about 0.01 $\mu_{\rm B}$ is left. The spin direction on apical O is opposite to that of Cu within the same CuO$_6$ octahedron. This small magnetic moment at apical O cannot be ignored in the estimation of the internal field at the muon site.\cite{Miyazaki} 

The Cu-spin density in the vertical direction to the CuO$_2$ plane was also investigated by DFT in order to visualize a magnetic path along the inter-plane direction within CuO$_6$ octahedron. However, almost no enlargement of the polarized spin-density distribution in Cu3$d_{z^2-r^2}$ was found as well as the preceding study,\cite{Lane} indicting a possibility that the inter-plane magnetic interaction could be driven by the direct exchange interaction between Cu3$d_{x^2-y^2}$ and 2$p_z$ of apical O within the same CuO$_6$ octahedron. On the other hand, the angle-resolved photoelectron spectroscopy succeeded to visualize Cu3$d_{z^2-r^2}$  and a $^{139}$La-NMR measurement pointed out a part of Cu-spin density was transferred to apical O via Cu3$d_{z^2-r^2}$.\cite{Matt,Watanabe} A theoretical study solving out low-energy Hamiltonians supposed active roles of Cu3$d_{z^2-r^2}$ hybridizing with O2$p_z$.\cite{Hirayama} These results indicate that the contribution of Cu3$d_{z^2-r^2}$ to the inter-plane magnetic interaction is not negligible and still an open question. 

\subsection{DFT+$U$ with $\mu$}
As the next step, DFT+$U$ was carried out including the muon to reproduce the $\mu$SR results. For this purpose, muon positions in LCO were investigated in advance. The precise determination of muon positions in LCO has not yet been successful in the past and was left as a long-term fundamental problem in the muon community.\cite{Hitti,Torikai,Sulaiman,Budi,Suter,PSI} 

\begin{figure}[htb]
\includegraphics[width=7cm]{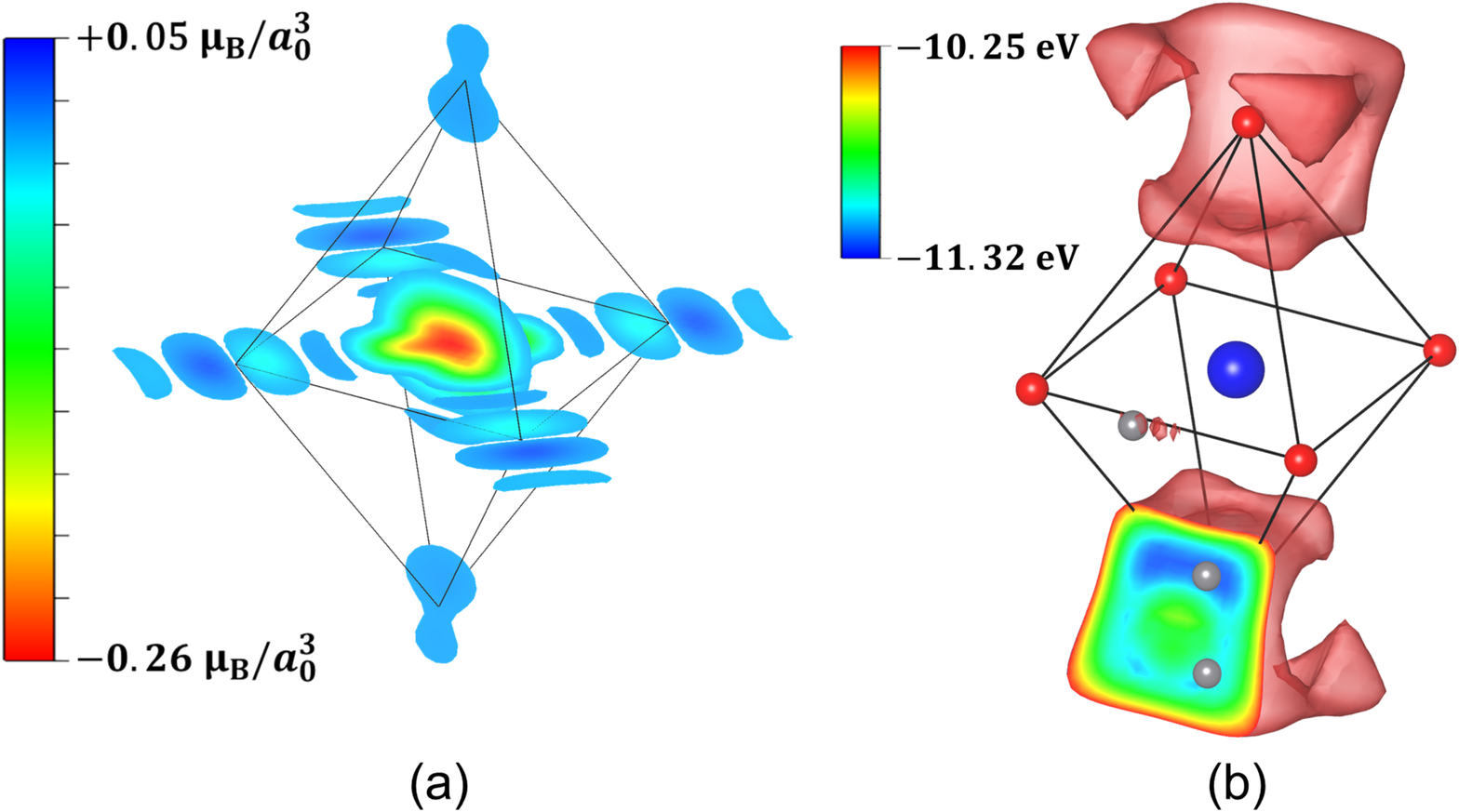}
\caption{(a) Three dimensional map of the Cu-spin density in the CuO$_6$ octahedron of La$_2$CuO$_4$ obtained from our DFT+$U$ calculations. The Cu-spin density expands to the in-plane O and reverses the spin direction at the other side centering the nuclear position of O as drawn by the  dark blue color. This means that the neighboring Cu-spin component which has the opposite spin direction as shown in Fig. \ref{spin_structure} flows into the same in-plane O and cancels the total spin component. The $a_0$ in the density unit is the Bohr radius.  (b) Electrostatic potential calculation results. The red area is the isosurface showing the energy level of 1.07 eV higher from the minimum potential. The energy level of the isosurface was chosen to make the position of M3$_{\rm DFT}$ clearly visible.  Gray balls indicate three local-minimum potential positions as candidates for initial muon stopped positions. }
\label{Cu_spin} 
\end{figure}

Figure \ref{Cu_spin}(b) shows our estimation of initial stopping positions of injected muons in LCO obtained from simple electrostatic potential calculations by using a unit cell since the muon has a positive charge and prefers to sit down at the minimum electrostatic potential just after it stops in the sample.\cite{Budi} Three possible local minimum potential positions were found as candidates of initial muon stopping positions. We named those three positions as M1$_{\rm DFT}$, M2$_{\rm DFT}$ and M3$_{\rm DFT}$ in order to compare with the $\mu$SR results. The injected muon choses one of those three positions to initially stop and moves to a local stable position interacting with surrounding atoms and electrons, causing local deformations of the crystal structure and electronic states in the vicinity of the muon.\cite{Suter,Moller}  There are four crystallographic equivalent sites of  each muon position within the unit cell. We confirmed by using the present calculation condition including the muon that those sites were also magnetically equivalent within the calculation accuracy.  

Since the number of injected muons is almost negligible compared with the number of atoms in the sample, the muon can be regarded as a super-dilute magnetic impurity in DFT+$U$. This situation requires us to set a sufficiently large supercell structure with one muon inside in order to follow the realistic $\mu$SR experimental condition. Accordingly, the 4$\times$4$\times$2 supercell was used for the present DFT+$U$ including the muon. DFT calculations on a small cell with one muon is unrealistic because the number of muons is comparable to the number of unit cell of the sample in smaller cells.\cite{Ramadhan} Our modeled supercell contained 896 atoms and one muon. All atomic positions and electronic density distributions were necessary to be as adjustable parameters within the supercell with only one muon as the magnetic impurity. This kind of supercell calculations need larger-scale computation volume. We carried out this large-scale calculation by using the high-performance supercomputing cluster system of HOKUSAI in RIKEN. 

\begin{figure}[htb]
\includegraphics[width=8cm]{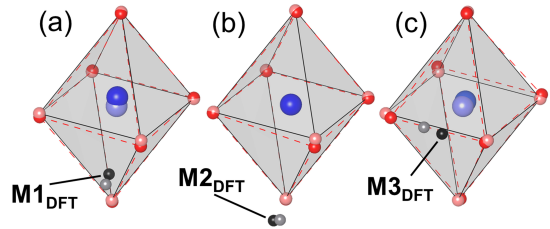}
\caption{Final stable muon positions for (a) M1$_{\rm DFT}$, (b) M2$_{\rm DFT}$ and (c) M3$_{\rm DFT}$, respectively, obtained from DFT+$U$ including the muon using the 4$\times$4$\times$2 supercell. Gray and black balls in each panel indicate the initial and final position of the muon after the relaxation, respectively.  }
\label{muon_position}
\end{figure}

Figure \ref{muon_position}(a)-(c) indicate the final muon positions and local deformations of the crystal structure estimated from our DFT+$U$+$\mu$ calculations. As for M1$_{\rm DFT}$, the muon is located near apical O and inside the CuO$_6$ octahedron as demonstrated in Fig. \ref{muon_position}(a). The muon moves into the more inside of the CuO$_6$ octahedron after relaxing its position and pushes away the Cu atom from the muon. Two in-plane O in the CuO$_2$ plane are pulled toward M1$_{\rm DFT}$. Concerning M2$_{\rm DFT}$, the muon stops near apical O as well as M1$_{\rm DFT}$ but outside the CuO$_6$ octahedron as shown in Fig. \ref{muon_position}(b). The muon moves away a little from apical O after the relaxation and pulls one in-plane O to its side. M2$_{\rm DFT}$ does not affect the Cu position too much. In terms of M3$_{\rm DFT}$, the muon sits in between two in-plane O and pulls them to its side and pushes away the Cu atom from the muon as exhibited in Fig. \ref{muon_position}(c). 

\begin{figure}[htb]
\includegraphics[width=8cm]{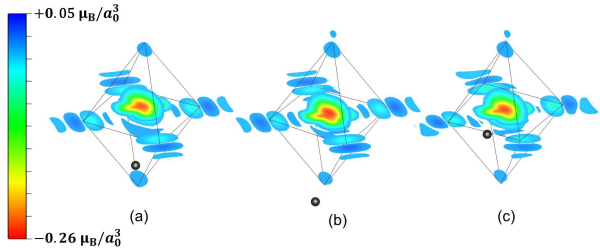}
\caption{Cu-spin density distribution estimated by DFT+$U$ including the muon using the 4$\times$4$\times$2 supercell with the muon at (a) M1$_{\rm DFT}$, (b) M2$_{\rm DFT}$ and (c) M3$_{\rm DFT}$, respectively. Black balls in each panel indicate the final position of the muon after the relaxation where the muon's density has the maximum. The $a_0$ in the density unit is the Bohr radius.}
\label{Cu_spin_muon} 
\end{figure}

Those local changes in atomic positions of Cu lead to changes in the Cu-spin distribution around the muon. Figure \ref{Cu_spin_muon} shows the results of DFT+$U$+$\mu$, indicating the Cu-spin density distribution around each muon position. In the case of M1$_{\rm DFT}$, the Cu-spin density becomes slightly less as shown in Fig. \ref{Cu_spin_muon}(a), resulting in the reduction of the magnetic moment of Cu in net. This reduction in the magnetic moment of Cu happens just beside the muon. The reduction ratio of the magnetic moment of Cu in the presence of the muon was estimated to be about -5 \% and -1 \% for M1$_{\rm DFT}$ and M3$_{\rm DFT}$, and $\sim$0 \% for M2$_{\rm DFT}$. 

The unbalanced spin density was also found at in-plane O which were caused by changes in the spin densities coming from adjacent two Cu atoms. This effect was the largest for M3$_{\rm DFT}$ as drawn in Fig. \ref{Cu_spin_muon}(c) and can be qualitatively understood as follows. The balance of the Cu-spin density transferred to in-plane O is broken due to deformations of local electronic states and crystal structure which happens at one side beside the muon. This unbalanced Cu-spin density at in-plane O causes the non-zero magnetic moment. The estimated size of the additional magnetic moment was about 0.01 $\mu_B$ but cannot be negligible for the estimation of the internal field at the muon site because this component appears just near the muon. Similarly, the small component of the magnetic moment at apical O slightly increased due to the change in the local Cu-spin density. All of these changes are local effects induced by the injected muon and disappear quickly beyond next neighbor unit cells.  

\subsection{Effects of the Zero-Point Vibration Motion of $\mu$}
In preceding studies, internal fields at the muon sites were always overestimated and could not explain the present $\mu$SR results even taking into account distributed Cu spins as shown in Fig. \ref{Cu_spin_muon}(a)-(c). Accordingly, we included one quantum effect of the muon itself, which was the zero-point vibration motion.\cite{Bernardini} This is because the muon is a fine particle with the lighter mass of about 1/9 compared to the hydrogen and has the spatial distribution around the stopping position following the shape of the local potential. This quantum motion of the muon can be obtained by solving the Schr\"odinger equation around the local potential surrounding the muon as follows. 
\begin{equation}
\left[ -\frac{\hbar^2 \nabla^2}{2 m_\mu} + V_\mu(r) \right] \psi_\mu(r) = E_\mu \psi_\mu(r).
\end{equation}
\noindent
Here, $V_\mu(r)$, $m_\mu$, $\psi_\mu(r)$ and $E_\mu$ are the potential around the muon, muon's mass, wave function and eigenvalue, respectively. The Schr\"odinger equation was solved out numerically by using the MATLAB program. 

\begin{figure}[htb]
\includegraphics[width=8cm]{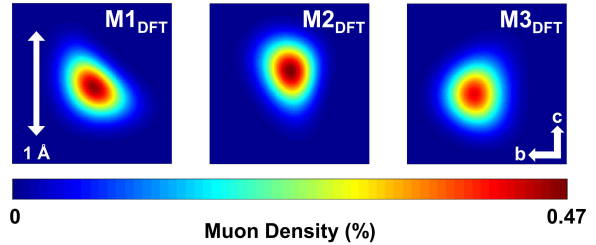}
\caption{Zero-point vibration motion of the muon itself around the minimum electrostatic potential for M1$_{\rm DFT}$, M2$_{\rm DFT}$ and M3$_{\rm DFT}$. These are views of cross sections perpendicular to the CuO$_2$ plain including the local minimum potential point.}
\label{zero_point} 
\end{figure}

Figure \ref{zero_point} shows the estimated muon-spin distribution due to the zero-point vibration motion around the minimum electrostatic potential for M1$_{\rm DFT}$, M2$_{\rm DFT}$ and M3$_{\rm DFT}$. These are two-dimensional images within the cross-section along the CuO$_2$ plane. Red- and blue-color regions indicate the high and low muon-density areas, respectively. We confirmed that more than 99 \% muon density are there within the 1.5 $\rm {\AA^3}$ cubic volume. The total sum of the dipole fields from surrounding Cu spins are calculated taking into account this muon-spin density distribution.

\section{Discussions}
In order to optimize muon positions and other related parameters, the internal fields at M1$_{\rm DFT}$, M2$_{\rm DFT}$ and M3$_{\rm DFT}$ were calculated on the basis of the dipole-dipole coupling, because LCO is a good insulator and the existing probability of conducting electrons around the muon is expected to be unlikely.\cite{Hyperfine} Cu-spin distributions and the muon's zero-point vibration motion were also included in the estimation of the internal field  by using the following equation.
\begin{equation}
\sum_{i,j} \frac{1}{|\vec{r}_i-\vec{r}_j|^3}
\bigg[3\vec{\rho_{i}}(\vec{r}_i-\vec{r}_j)
\frac{(\vec{r}_i-\vec{r}_j)}{|\vec{r}_i-\vec{r}_j|^2} - \vec{\rho_{i}}\bigg]|\psi_{j}|^2.
\label{dipole}
\end{equation}
\noindent
Here, $\vec{\rho_{i}}$ is the vector data for the spin grids and $\vec{r}_i-\vec{r}_j$ is the relative distance between the Cu-spin-density grids with the density of $\rho_{i}$ and the muon probability grids $|\psi_{j}|^2$. Then, we summed up all grid components obtained from our DFT calculations to estimate the internal field at the muon site. We set the radius of 50 \AA\ centering the muon to achieve the converged results for dipole calculations. The supercell with one muon was set at the center of the calculated sphere and other areas were filled up by normal unit cells without muons. 

It should be noted that multiple magnetic sites can be realized when different spin-structure domains are induced around microscopic defects in the LCO crystal as suggested from the $\mu$SR study on the LCO thin film.\cite{PSI} If this is the case, those defects are expected to introduce free Cu spins as well around defects in the Cu-spin network.\cite{Kojima2,Adachi} Our magnetic susceptibility measurement on the LCO single crystal showed that the fraction of those kinds of free spins is very small and almost negligible. This result means that the LCO single crystal used in the present study has less defects, indicating the uniform Cu-spin network with a single spin-structure domain. 

We have already simulated internal fields at three muon sites in LCO with some different spin structures including the ones suggested from the $\mu$SR study on the LCO thin film.\cite{Ramadhan2} This previous result showed that three different internal fields at our estimated three muon sites in the LCO single crystal can be quantitatively explained by one magnetic domain even though different spin states were set. Accordingly, we assumed in the present study that one uniform magnetic spin structure which was the same as that determined from the neutron scattering experiment appeared in the LCO single crystal,\cite{Budnick,Vaknin} resulting in that the existence of three muon sites in LCO was intrinsic. 

Since all muon positions, magnetic moment of Cu and Cu-spin density distributions are related to $U$, the dipole-field calculation was repeated varying $U$ from 2 to 8 eV in order to find out the optimized results. All calculated values which we have done are summarized in Table \ref{Diople_Results}. 

\onecolumngrid

\begin{center}
%\vspace{-1.1cm}
\begin{table}[htb]
\fontsize{9pt}{0.5cm}\selectfont
\caption{Top part: Magnetic moment of the Cu spin estimated from current DFT calculations without the muon by varying $U$. Bottom part: Calculated internal fields at M1$_{\tiny \rm {DFT}}$, M2$_{\tiny \rm {DFT}}$ and M3$_{\tiny \rm {DFT}}$ obtained from DFT calculations by varying $U$. All calculations were done with the same conditions taking into account the local deformation of the crystal structure and electronic state lead by the muon. The zero-point vibration motion of the muon was also included in the calculation. } 
\vspace{0.5cm}
\begin{tabular}{c||ccccccccccccc} \hline
            & \multicolumn{13}{c}{$U$ (eV)} \rule[0mm]{0mm}{3mm}\\ \cline{2-14}
            &2 & 3 & 3.5 & 4 & 4.5 & 5 & 5.6 & 6 & 6.5 & 7 & 7.2 & 7.5 & 8  \rule[0mm]{0mm}{3mm} \\ \hline \hline
     & \multicolumn{13}{c}{Calculated Magnetic Moment without $\mu$ ($\mu_{\rm B}$)} \rule[0mm]{0mm}{3mm}\\ \hline
Magnetic Moment & 0.386 & 0.436 & 0.460 & 0.482 & 0.503 & 0.524 & 0.547 & 0.562 & 0.583 & 0.602 & 0.609 & 0.621 & 0.641  \\ \hline \hline
{Muon Position} & \multicolumn{13}{c}{Calculated Internal Fields, $H^{\rm M \it i}_{\rm DFT}$ (G)} \rule[0mm]{0mm}{3mm}\\ \hline
M1$_{\tiny \rm {DFT}}$ & 336.35 & 376.01 & 384.85 & 391.51 & 429.28 & 446.36 & 450.22 & 471.50 & 474.90 & 491.08 & 503.83 & 507.16 & 523.10 \\
M2$_{\tiny \rm {DFT}}$ & 103.99 & 116.30 & 121.28 & 127.04 & 131.21 & 135.68 & 141.54 & 145.85 & 150.14 & 154.33 & 155.44 & 159.07 & 163.76  \\
M3$_{\tiny \rm {DFT}}$ & 888.06 & 990.60 & 1038.67 & 1077.89 & 1128.28 & 1169.74 & 1219.72 & 1241.02 & 1281.64 & 1322.98 & 1334.41 & 1364.36 & 1407.01\\ \hline 
\end{tabular}
\label{Diople_Results}
\end{table}
\end{center}

\twocolumngrid

\noindent
Basised on those results, we indexed the calculated internal fields as $H^{\rm M \it i}_{\rm DFT}$ ($i$=1,2,3) in order to compare with $H^{\rm M \it i}_{\mu {\rm SR}}$. And then, we defined differences between both values by using the following equation.  

\begin{equation}
\Delta H_{\rm M \it i} = (H^{\rm M \it i}_{\rm DFT} - H^{\rm M \it i}_{\mu {\rm SR}}) , (i = 1,2,3).
\label{difference}
\end{equation}

\noindent
After this, we summed up all $\Delta H_{\rm M \it i}$ for each $U$ with fitting-error values of internal fields, $\sigma_{\it i}$, as follows. Detail values of $\sigma_{\it i}$ are listed in Table \ref{Analysis_Results}. 

\begin{equation}
\sum_{i} \frac{\Delta H_{\rm M \it i}^2}{\sigma_{\it i}^2} , (i = 1,2,3).
\label{sum_difference}
\end{equation} 

\begin{figure}[t]
\includegraphics[width=6cm]{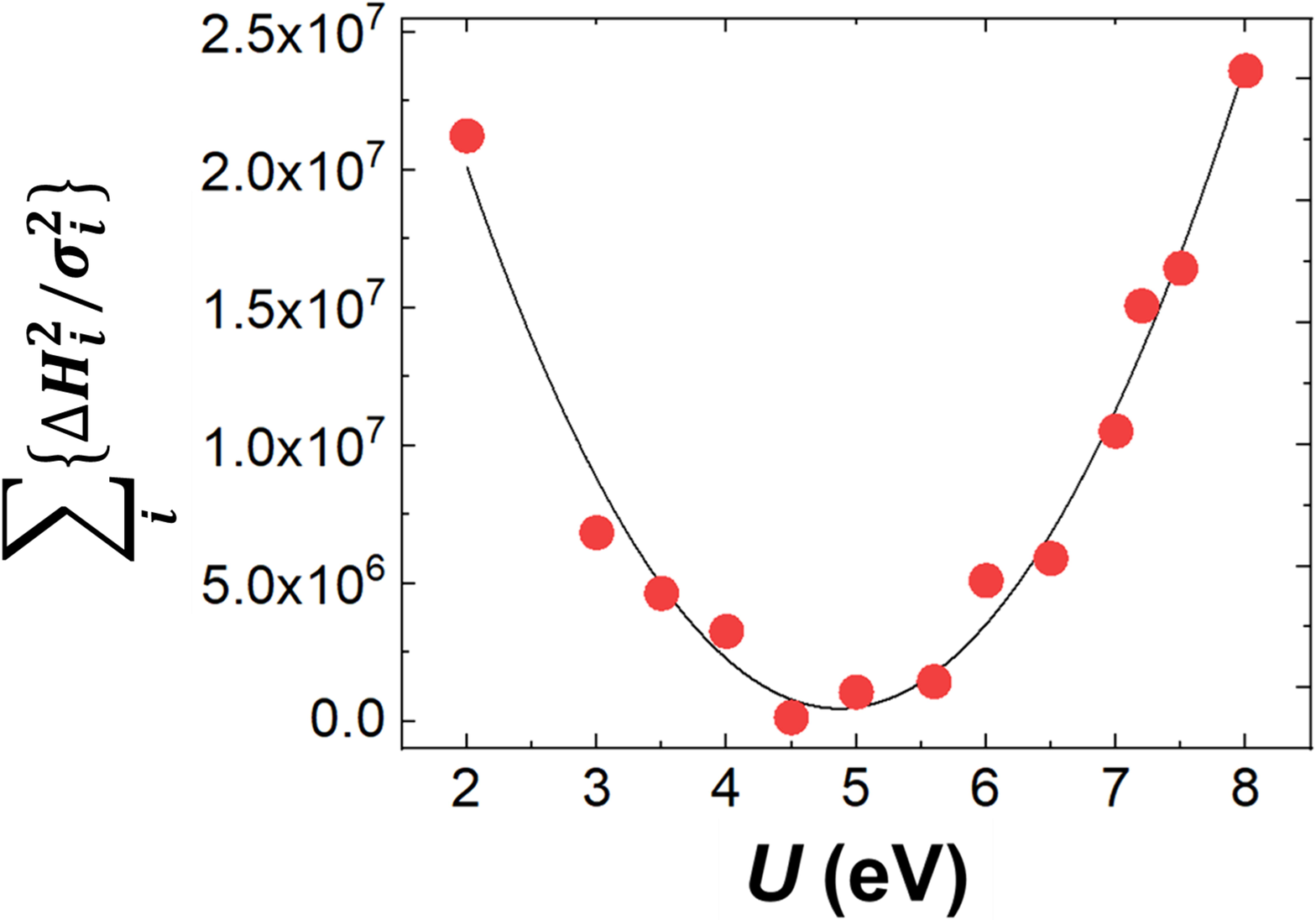}
\caption{Optimization of $U$ in terms of the difference in internal fields obtained by $\mu$SR and DFT+$U$+$\mu$ calculations, varying $U$ from 2 to 8 eV. The solid line is the best-fit result by using the Gaussian function. $\Delta H_{\rm M \it i}$ is difference between $H^{\rm M \it i}_{\rm {\mu SR}}$  and $H^{\rm M \it i}_{\rm DFT}$ ($i$=1,2,3) as described in Eq. \ref{difference}.  $\sigma_{\it i}$ is the fitting-error value of the internal field  at each $H^{\rm M \it i}_{\mu {\rm SR}}$.   }
\label{U_fit}
\end{figure}

\noindent
Figure \ref{U_fit} shows the $U$ dependence of summed up values obtained from Eq. (\ref{sum_difference}). Applying the Gaussian function, $U$ was optimized to be 4.87(4) eV. This value locates at the lower-end of the $U$ range which has been argues to be from 3 to 10 eV.\cite{Czyzyk,Anisimov0,Wan,Pesant,Werner,Jang,Hirayama,Lane,Nilsson} 

It was pointed out from theoretical studies on Hubbard model with a square lattice that $U$ is strongly correlated to the energy scale of the effective spin Hamiltonian which is described as 4$t^2$/$U$ and that a border between strong and weak correlations is around $U$$\sim$6.5$t$.\cite{Ogata} Here, $t$ is the hopping energy of electrons. Our present result of the smaller $U$ in LCO would give limitations on discussions of $t$ and provide possible dedicated directions to understand differences in $T_{\rm c}$ among high-$T_{\rm c}$ superconducting cuprates.\cite{Zaanen} For instance, we suggest following the {\it ab-initio} calculation of the effective Hamiltonian that the electronic state of LCO is closer to the one-band model although more detail comparisons with theoretical investigations are necessary.\cite{Hirayama} 

Following this result of the optimization of $U$, the magnetic moment of Cu was estimated through the same DFT calculation processes. The optimized value was 0.520(3) $\mu_B$ in the case of LCO without the muon. The direction of the optimized Cu spin was still along the b-axis after the non-collinear refinement. The estimated magnetic moment and the spin structure were consistent with those suggested from neutron scattering experiment.\cite{Vaknin}  In addition, optimized internal fields at each muon site in the case of $U$ = 4.87(4) eV were calculated from the dipole-field calculation using Eq.(\ref{dipole}) to be 429.7(12) G,  134.1(4) G and 1,147.6(35) G for M1$_{\rm DFT}$, M2$_{\rm DFT}$ and M3$_{\rm DFT}$, resulting in that M1$_{\rm DFT}$ = M1$_{\mu}$, M2$_{\rm DFT}$ = M2$_{\mu}$ and M3$_{\rm DFT}$ = M3$_{\mu}$, respectively. Differences in the internal field between the $\mu$SR and DFT+$U$+$\mu$ were 3.4 G ($\sim$1\%), 38.2 G ($\sim$40 \%) and 97.9 G ($\sim$8 \%) for M1$_{\mu}$, M2$_{\mu}$ and M3$_{\mu}$, respectively. All optimized muon positions and internal fields at there are summarized in Table \ref{all_fields}. Atomic positions of the CuO$_6$ octahedron after the opitmization with the muon at M1$_{\rm DFT}$, M2$_{\rm DFT}$ and M3$_{\rm DFT}$ are listed in Table \ref{all_positions}.   

\begin{figure}[t]
\includegraphics[width=6cm]{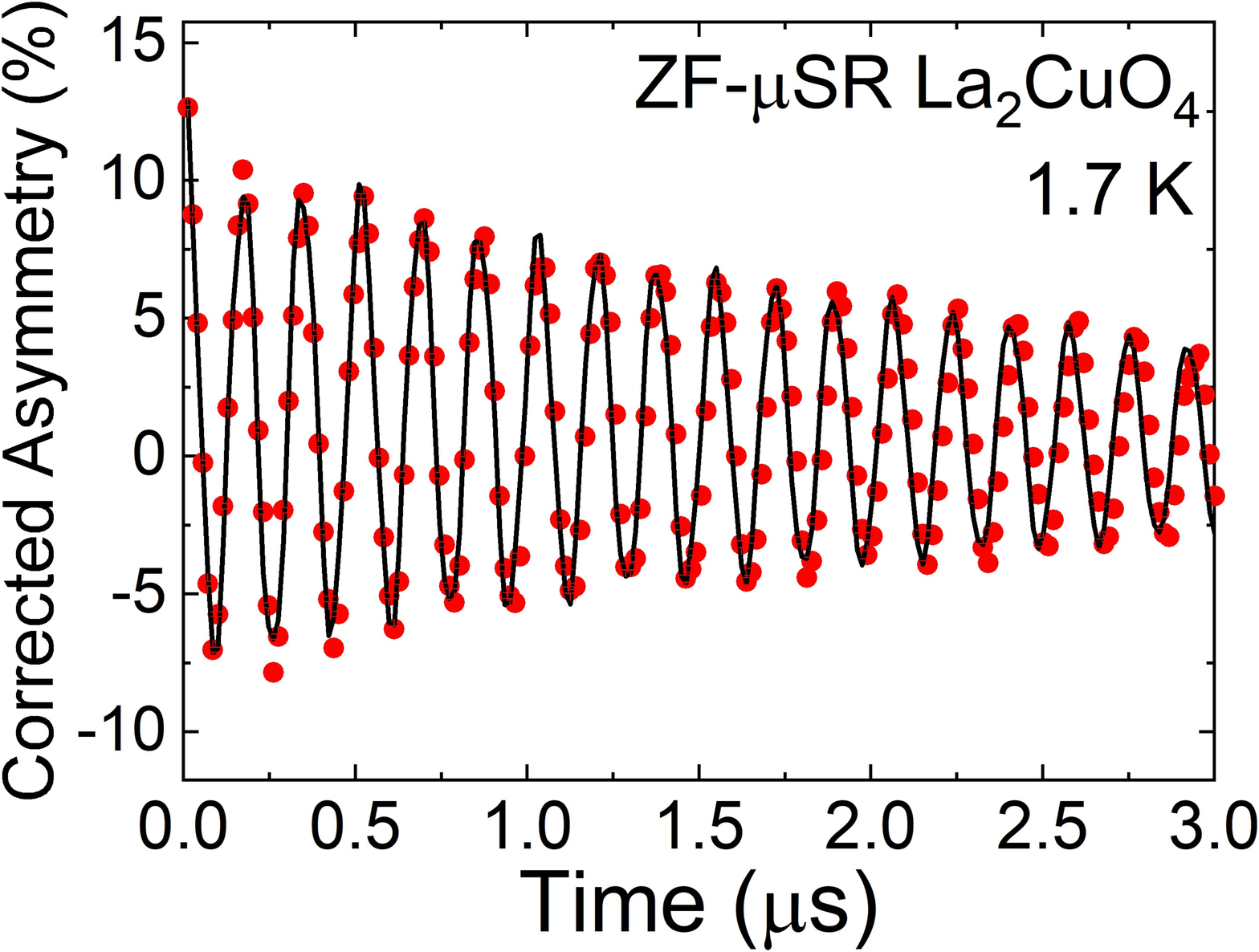}
\caption{ZF-$\mu$SR time spectrum observed at 1.7 K with the simulated line by using internal fields which were estimated from the current DFT calculations. The black-solid line is the trace of the simulation. Same values for $A_i$, $\phi_i$ and $\lambda_i$ listed in Table \ref{Analysis_Results} were used, replacing $H^{\rm M \it i}_{\rm {\mu SR}}$ to be $H^{\rm M \it i}_{\rm DFT}$ to draw the black-solid line.   }
\label{spectrum_simulation}
\end{figure}

Figure \ref{spectrum_simulation} indicates the simulated $\mu$SR time spectrum by using internal fields obtained from present DFT+$U$+$\mu$ calculations. We used the same values for $A_i$, $\phi_i$ and $\lambda_i$ as listed in Table \ref{Analysis_Results} and $H^{\rm M \it i}_{\rm DFT}$ in order to evaluate our DFT results. The solid-red line in Fig. \ref{spectrum_simulation} is the simulation result. The simulated result reproduced the time spectrum fairly well, but there were still small differences between measured and simulated $\mu$SR time spectra, especially in the longer time region. This is because, simulated internal fields for M1 and M3 are very close to the experimental results but the one for M2 is still fairly far. The reason why our DFT+$U$+$\mu$ did not perfectly reproduce the experimental result is guessed to be due to DFT's underlying principal statistical errors with regards to the pseudo-potential approximation, calculation-grid resolution, cut-off energy, relaxation step for self-consistent calculation loop and so on. Although the DFT+$U$ calculation has been well established to describe electronic states of strongly correlated systems\cite{Czyzyk,Anisimov0,Wan,Pesant,Werner,Jang,Hirayama,Lane,Nilsson,Varignon,Moller} and those statistical errors should be small, errors would be piled up during the total-energy minimization process of the non-periodical supercell model with the muon and become non-ignorable as a result in our case. 

\onecolumngrid

\begin{table}[h]
\fontsize{10pt}{0.5cm}\selectfont
\caption{Cartesian components of optimized muon positions in the 4$\times$4$\times$2 supercell and internal fields at each muon positions in the style of normalized component against the unit cell size along $a$-, $b$- and $c$-axis. The definition of each crystal axis was the same with that used in the neutron scattering experiment.\cite{Vaknin} The negative signature means that internal fields direct opposite. } 
\vspace{0.5cm}
\begin{tabular}{c||ccc|ccc||ccccc} \hline
\multirow{2}{*}{Muon Position} 
&\multicolumn{3}{c|}{before relaxation}&\multicolumn{3}{c||}{after relaxation}& \multicolumn{5}{c}{Internal Fields (G)}  \rule[0mm]{0mm}{3mm}\\ \cline{2-4} \cline{5-7} \cline{8-12}
& $a$ & $b$ & $c$ & $a$ & $b$ & $c$ & Fourier & DFT & $a$ & $b$ & $c$ \rule[0mm]{0mm}{3mm} \\ \hline \hline
M1$_{\tiny \rm {DFT}}$ & 0.3839 & 0.5982 & 0.4336 & 0.3777 & 0.6175 & 0.4375 & 426.59(1) & 429.7(12) & -2.81 & 355.06 & -241.92\\
M2$_{\tiny \rm {DFT}}$ & 0.3928 & 0.5893 & 0.4023 & 0.3817 & 0.5968 & 0.3975 & 109.16(39)& 134.1(4) & -8.54 & 73.89 & -111.04\\
M3$_{\tiny \rm {DFT}}$ & 0.3660 & 0.5491 & 0.4961 & 0.3880 & 0.5502 & 0.4935 & 1251.55(27)& 1147.6(35)& 23.47 & -1134.59 & -170.17 \\ \hline
\end{tabular}
\label{all_fields}
\end{table}

\begin{table}[h]
\fontsize{10pt}{0.5cm}\selectfont
\caption{Cartesian components of atomic positions in the 4$\times$4$\times$2 supercell before and after the opimization of the CuO$_6$ octahedron with the injected muon at M1$_{\tiny \rm {DFT}}$, M2$_{\tiny \rm {DFT}}$ and M3$_{\tiny \rm {DFT}}$, respectively. Each atomic position in CuO$_6$ octahedron is indicated in the figure at the right-end of the table. All positions are described in the style of normalized component against the unit cell size along $a$-, $b$- and $c$-axis. The definition of each crystal axis was the same with that used in the neutron scattering experiment.\cite{Vaknin}   } 
%\hspace{-3cm}
\begin{tabular}{c||ccc|ccc|ccc|ccc|c} \hline
 & \multicolumn{3}{c|}{before relaxation}&\multicolumn{9}{c|}{after relaxation}  & \multirow{9}{*}{
\begin{minipage}{30mm}
  \centering
  \scalebox{0.018}[0.018]{\includegraphics{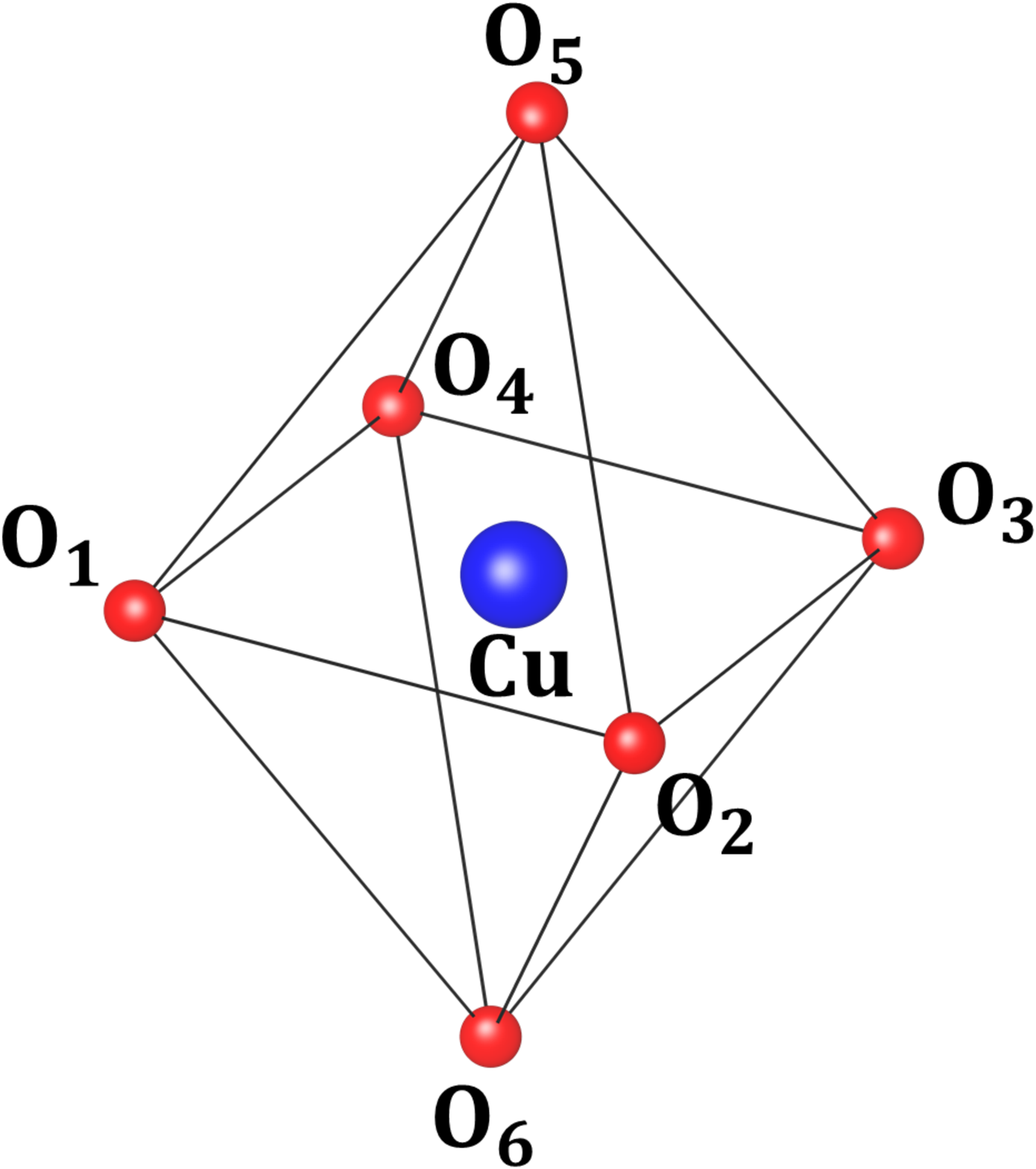}}
\end{minipage}
}  \rule[0mm]{0mm}{3mm}\\  \cline{2-13}  
 & \multicolumn{3}{c|}{without $\mu$}    &\multicolumn{3}{c|}{with $\mu$ at M1$_{\tiny \rm {DFT}}$} &\multicolumn{3}{c|}{with $\mu$ at M2$_{\tiny \rm {DFT}}$} & \multicolumn{3}{c|}{with $\mu$ at M3$_{\tiny \rm {DFT}}$} &  \rule[0mm]{0mm}{3mm}\\  \cline{2-13}  
{Atoms} &  $a$ & $b$ & $c$ & $a$ & $b$ & $c$ & $a$ & $b$ & $c$ & $a$ & $b$ & $c$  & \rule[0mm]{0mm}{3mm} \\ \cline{1-13}  \cline{1-13}
   Cu       & 0.3750 & 0.6250 & 0.5000 & 0.3751 & 0.6243 & 0.5087 & 0.3757 & 0.6240 & 0.5003 & 0.3739 & 0.6322 & 0.5024  & \\
   O$_1$ & 0.3125 & 0.5625 & 0.4964 & 0.3128 & 0.5635 & 0.4915 & 0.3125 & 0.5620 & 0.4942 & 0.3202 & 0.5597 & 0.4932  & \\
   O$_2$ & 0.4375 & 0.5625 & 0.4964 & 0.4369 & 0.5639 & 0.4891 & 0.4377 & 0.5622 & 0.4874 & 0.4350 & 0.5623 & 0.4929 &  \\
   O$_3$ & 0.4375 & 0.6875 & 0.5036 & 0.4373 & 0.6876 & 0.5056 & 0.4381 & 0.6869 & 0.5042 & 0.4376 & 0.6898 & 0.5084  & \\
   O$_4$ & 0.3125 & 0.6875 & 0.5036 & 0.3128 & 0.6876 & 0.5059 & 0.3133 & 0.6872 & 0.5043 & 0.3102 & 0.6891 & 0.5087  & \\
   O$_5$ & 0.3750 & 0.6165 & 0.5918 & 0.3755 & 0.6112 & 0.5917 & 0.3766 & 0.6122 & 0.5918 & 0.3762 & 0.6079 & 0.5927  & \\
   O$_6$ & 0.3750 & 0.6335 & 0.4018 & 0.3742 & 0.6402 & 0.4052 & 0.3733 & 0.6408 & 0.4057 & 0.3753 & 0.6418 & 0.4076  & \\  \hline
\end{tabular}
\label{all_positions}
\end{table}

\twocolumngrid

\vspace{5cm}

It is worthwhile to describe other value-added results obtained from the present DFT study. By using the optimized $U$, the band-gap structure can also be optimized, leading to the minimum CT gap between the upper Hubbard band and O2$p$ to be 1.24(1) eV. This CT-gap value has been discussed within the range of 0.9-2 eV giving large ambiguity.\cite{Lane,Tokura,Uchida,Ohno} Note that our obtained value is in the ground state at 0 K. Even taking into account that the measured CT gap shows a shift for a couple of 0.1 eV to the lower energy side with increasing temperature,\cite{Ohno} our obtained value is fully consistent with the previous results.\cite{Lane,Tokura,Uchida,Ohno} Those facts also proved that our results revealed the realistic feature of the electronic state of LCO.  

There still be one more question left for the full understandings of the $\mu$SR results. That is how to explain differences in populations of stopped muons among the three sites. The experimental results indicate that most of injected muons stop at M1$_{\mu}$ as evidenced in Fig. \ref{time_spectrum}(b). The ratio of populations of muons among those three sites were determined from the differences in the initial asymmetries to be as M1$_{\mu}$:M2$_{\mu}$:M3$_{\mu}$=106:30:10. One possible way to address this question is to model the stopping procedure of the muon in LCO after its injection. This is left as an open question. More DFT calculations and/or simulations will be required to tackle this problem. 

\section{Conclusion}
We determined the value of $U$, covalent state of the Cu spin and the CT gap energy in LCO by combining $\mu$SR experiments and DFT calculations. Three muon positions in LCO were identified and $U$ was precisely determined to be 4.87(4) eV, followed by the magnetic moment of Cu to be 0.520(3) $\mu_{\rm B}$ and the minimum CT gap between the upper Hubbard band to the O2$p$ band to be 1.24(1) eV. The role of the perturbation introduced by the muon was found to deform the local crystal structure just around the muon, followed by subsequent changes in the surrounding electronic state in LCO. This effect leads to the slight reduction in the magnetic moment surrounding the muon. 

Strong benefit of our technique is that we can achieve information of the spin structure, size of magnetic moment, muon positions and $U$ in one time by analyzing one $\mu$SR time spectra.  Especially, the $U$ value cannot be optimized from other experimental methods with good accuracy as demonstrated in the present study. In addition, our technique is workable for other systems on the basis of some experimental and computational conditions. Those are; 1) the target system has magnetic moments, 2) the muon-spin precession should be observed and 3) DFT calculations is applicable, 4) there are accessible high-performance computing resources which can accept large-scale supercell calculations. As long as those four conditions are satisfied, our developed technique to estimate $U$ is widely applicable to any systems. For instance, mother systems of all Cu-based high-$T_{\rm c}$ cuprates, Mott systems, heavy Fermions and strongly correlated organic molecular systems are good targets. Even using other DFT package programs like Quantum Espresso, CASTEP and Wien2K, one can apply the same method described in this report to their own target materials. This means that the transferability of our method to other materials is quite high and widely applicable to other research fields providing us deeper knowledge on their unique and exotic properties from a different perspective via $\mu$SR.

\begin{acknowledgments}
The authors would like to thank for technical supports by the muon group of PSI to carry out the $\mu$SR measurement and also thank K. Ishida, A. Fujimori and M. Ogata for their valuable discussions. We would like to acknowledge the HOKUSAI supercomputing facility (Project. No. G19007) of RIKEN. This work is supported by JSPS KAKENHI (No's JP19H01841 and 20H04463) and International Program Associate of RIKEN.
\end{acknowledgments}

% Create the reference section using BibTeX:
\bibliography{Ramadhan-DFT-Ref}

\end{document}